\documentclass
[prl,twocolumn,superscriptaddress,floatfix,showpacs,10pt]{revtex4}%
\usepackage{amsfonts}
\usepackage{amsmath}
\usepackage{amssymb}
\usepackage{mathrsfs}
\usepackage{xcolor}
\usepackage{graphicx}%
\usepackage[normalem]{ulem} 
\font\mybb=msbm10 at 10pt
\def\bb#1{\hbox{\mybb#1}}

\def\bR {\bb{R}}

\def\ch{{\rm ch\,}}
\def\sh{{\rm sh\,}}

\begin{document}

\title{A non-linear duality-invariant  conformal extension of Maxwell's equations}

\author{Igor Bandos}
%\email{igor.bandos@ehu.eus}
\affiliation{Department of Theoretical Physics, University of the Basque Country UPV/EHU, 
P.O. Box 644, 48080 Bilbao, Spain, and  IKERBASQUE, Basque Foundation for Science, 
48011, Bilbao, Spain}

\author{Kurt Lechner}
%\email{kurt.lechner@pd.infn.it}
\author{Dmitri Sorokin}
%\email{dmitri.sorokin@pd.infn.it}
\affiliation{I.N.F.N., Sezione di Padova, 
and  Dipartimento di Fisica e Astronomia “Galileo Galilei”, 
Universit\'a degli Studi di Padova, 
Via F. Marzolo 8, 35131 Padova, Italy}

 \author{Paul K. Townsend}
%\email{p.k.townsend@damtp.cam.ac.uk}
\affiliation{Department of Applied Mathematics and Theoretical Physics, Centre for
Mathematical Sciences, University of Cambridge, Wilberforce Road, Cambridge,
CB3 0WA, UK}

\begin{titlepage}
\vfill
%\centerline{\Large{ DRAFT}}
%\vfill
\end{titlepage}

\begin{abstract}

All nonlinear extensions of the source-free Maxwell equations preserving both $SO(2)$ electromagnetic duality invariance and conformal invariance are found, and shown to be 
limits of a one-parameter generalisation of  Born-Infeld electrodynamics. The strong-field limit 
is the same as that found by  Bialynicki-Birula from Born-Infeld theory but the weak-field limit is a new one-parameter extension of Maxwell electrodynamics, which is interacting but admits 
exact light-velocity plane-wave solutions of arbitrary polarisation. Small-amplitude waves on a constant uniform electromagnetic background exhibit birefringence, but one polarisation 
mode remains lightlike.

\end{abstract}

\maketitle

\setcounter{equation}{0}
  
  Nonlinear extensions of Maxwell's equations have a long history. The Born-Infeld (BI) equations \cite{Born:1934gh}, which preserve the electromagnetic duality invariance of Maxwell's 
equations, is perhaps the best known example, for reasons reviewed in \cite{Gibbons:2001gy}. The Euler-Heisenberg equations \cite{Heisenberg:1935qt}, which incorporate vacuum 
polarisation effects of QED,  is another. Both reduce to  Maxwell's equations in a weak-field limit since the interactions introduce a fixed energy scale that breaks conformal invariance. 

It is to be expected that any nonlinear electrodynamics theory in a Minkowski spacetime will have some conformal 
weak-field limit, and it is easily established that this must be Maxwell electrodynamics if it is assumed that all conformal invariant equations arise as Euler-Lagrange (EL) equations for a  
Lagrangian density that is a real analytic function of gauge-invariant Lorentz  scalars constructed from electric and magnetic fields only.  However, this assumption
also implies that the only possible conformal strong-field limit is Maxwell's theory, whereas the  strong-field limit of  BI theory has long been known 
to be an interacting conformal theory  \cite{BialynickiBirula:1984tx,BialynickiBirula:1992qj},  which we shall call  Bialynicki-Birula (BB) electrodynamics.  

Here we use Hamiltonian methods similar to those of \cite{BialynickiBirula:1984tx} to investigate whether there are other physically acceptable interacting conformal 
electrodynamics theories. We restrict our analysis to theories with an electromagnetic duality invariance, for simplicity but also because of its intrinsic interest; one of 
our aims is to determine whether there are interacting alternatives to Maxwell electrodynamics with the same symmetries. 
We find that the conditions imposed by duality and conformal invariance on the Hamiltonian density of a generic nonlinear theory of electrodynamics have two types of 
solution: one yields BB electrodynamics and the other yields a generalisation of Maxwell electrodynamics, which we call the ModMax theory. The  ModMax equations 
depend on a dimensionless parameter $\gamma$, and they reduce to Maxwell's equations  for $\gamma=0$. For any other value of $\gamma$ the equations are nonlinear,  
but for $\gamma\ge0$ they admit exact light-velocity plane wave solutions of arbitrary polarisation. 

The Lagrangian formulation of ModMax electrodynamics is found by a Legendre transform. As expected, the Lagrangian density is not analytic everywhere; it fails to be analytic at 
configurations for which the Lorentz invariants are zero, which includes the vacuum.  Linearisation about a non-vacuum background of uniform constant electromagnetic fields is possible, however,  
and it leads to a polarisation-dependent dispersion relation for small-amplitude waves, i.e. birefringence, as expected since BI theory is known to be the unique nonlinear electrodynamics that does not exhibit  birefringence \cite{Boillat:1970gw,Plebanski:1970zz}. For the  ModMax theory  we shall see that there is always one light-velocity polarisation mode, while the other  is subluminal for 
$\gamma>0$ but superluminal for $\gamma<0$. This provides a physical reason for a restriction to $\gamma\ge0$. It also has implications for predictions derived from the
Euler-Heisenberg theory, as we explain in our conclusions. 

The Hamiltonian density ${\cal H}$ for a generic source-free theory of electrodynamics is a function of the magnetic induction $3$-vector field ${\bf B}$ and an independent electric-displacement $3$-vector field ${\bf D}$.   The field equations are the ``macroscopic Maxwell equations''
\begin{eqnarray}\label{feqs}
\dot {\bf B}  &=&   -\boldsymbol{\nabla}\times {\bf E} \, , \qquad \boldsymbol{\nabla} \cdot {\bf B} =0\, , \nonumber\\
\dot {\bf D}  &=& {}\;\; \boldsymbol{\nabla} \times {\bf H} \, , \qquad  \boldsymbol{\nabla} \cdot {\bf D} =0\, , 
\end{eqnarray}
taken together with the ``constitutive relations'' 
\begin{equation}\label{EHdefs}
{\bf E} =  \partial {\cal H}/\partial {\bf D} \, , 
\qquad {\bf H} = \partial {\cal H}/\partial {\bf B} \, .
 \end{equation}
These equations imply that 
 \begin{equation}\label{continuity}
\dot {\cal H}=  -\boldsymbol{\nabla} \cdot ({\bf E}\times{\bf H}) \, , \qquad  \dot {\cal P}_i = -\partial_j T^j{}_i\, , 
\end{equation}
where $\{{\cal P}_i\, ; i=1,2,3\}$ are the components of the field $3$-momentum density $\boldsymbol{\cal P} =  {\bf D} \times{\bf B}$, and 
\begin{equation}
T^i{}_j  = \delta^i_j  \left( {\bf B}\cdot{\bf H} + {\bf D}\cdot {\bf E} - {\cal H}\right) - \left(B^iH_j + D^iE_j\right)\, . 
\end{equation}
This is the stress tensor; it is symmetric because rotational invariance implies that ${\bf B}\times {\bf H}+ {\bf D}\times{\bf E} ={\bf 0}$. 

We may conclude from (\ref{continuity}) that the  integrals over space of ${\cal H}$ and $\boldsymbol{\cal P}$ are conserved quantities for appropriate 
boundary conditions; they are the  conserved energy and momentum associated with the time and space translational invariance of the 
field equations. Together with rotational invariance, these are the manifest symmetries of the field equations but there may be additional 
symmetries that are not manifest, such as Lorentz boost invariance. In a Lorentz invariant theory it should be possible to write the equations (\ref{continuity}) as the $4$-vector continuity equation
for a {\it symmetric}  stress-energy  tensor, but this is possible only if 
\begin{equation}\label{BBLI}
{\bf E} \times {\bf H} = {\bf D} \times {\bf B}\, , 
\end{equation} 
which is therefore the condition for the equations (\ref{feqs}) to be Lorentz invariant \cite{BialynickiBirula:1984tx}.  The  Lorentz scalar trace of this stress-energy  4-tensor is 
\begin{equation}
T^i{}_i - {\cal H} = 2 \left[ {\bf D}\cdot {\bf E} + {\bf B} \cdot{\bf H} -2{\cal H}\right] \, .  
\end{equation}
The condition for conformal invariance is therefore  (\ref{BBLI}) {\it and} 
\begin{equation}\label{conf-inv} 
{\bf D}\cdot {\bf E} + {\bf B}\cdot {\bf H} = 2{\cal H}\, . 
\end{equation} 
Finally, the condition for invariance under the $SO(2)$ electromagnetic duality group,  which acts by shifting the phase of the complex 3-vector field $ {\bf D} +i{\bf B}$,  
is \cite{BialynickiBirula:1984tx}
\begin{equation}
{\bf E} \cdot{\bf B} = {\bf D}\cdot {\bf H}\, . 
\end{equation}

There are three independent rotation scalars, but at most two are  duality invariant; for example:
\begin{equation}
s= \frac12( |{\bf D}|^2 + |{\bf B}|^2)\, , \qquad  p = |{\bf D}\times {\bf B}| \, .  
\end{equation} 
If ${\cal H}$ is a duality invariant rotation scalar, which we assume,  then it must be a function of $s$ and $p$. Using the notation $({\cal H}_s,{\cal H}_p)$ for partial derivatives of ${\cal H}$, 
the Lorentz invariance condition (\ref{BBLI}) implies, upon using (\ref{EHdefs}), that
\begin{equation}\label{PDEsp}
{\cal H}^2_s + \frac{2s}p{\cal H}_s{\cal H}_p + {\cal H}^2_p  =1 \, . 
\end{equation}
A convenient alternative basis for the duality-invariant rotation scalars is
\begin{equation}\label{BBvars}
{\tt u} = \frac12(s+ \sqrt{s^2-p^2}) \, , \quad {\tt v} = \frac12(s- \sqrt{s^2-p^2})\, .   
\end{equation}
These new variables are well-defined since 
\begin{equation}\label{xieta}
s^2-p^2 =  \xi^2 + \eta^2 \ge0\, , 
\end{equation}
where $(\xi,\eta)$ are the following rotation scalars (which will be used later):
\begin{equation}\label{newrot}
\xi = \frac12(|{\bf D}|^2-|{\bf B}|^2) \, , 
\qquad  \eta={\bf D}\cdot {\bf B} \, .  
\end{equation}

If we look for solutions of the form ${\cal H} = \sqrt{K} + {\rm constant}$, then the equation we need to solve is 
$K_{\tt u} K_{\tt v} =4K$.  For quadratic $K$ this just restricts the coefficients, and the general quadratic solution for which $K$ is non-negative for all
$({\tt u},{\tt v})$ is found to depend on one parameter $T$ with dimensions of energy density and an additional dimensionless
parameter $\gamma$ such that,  for zero vacuum energy, 
\begin{equation}\label{newHam} 
{\cal H} = \sqrt{ T^2 + 2T(e^{-\gamma} {\tt u} + e^{\gamma} {\tt v}) + 4 {\tt u}{\tt v}} -T \, . 
\end{equation}
This is the Hamiltonian density of BI electrodynamics when $\gamma=0$. The strong-field ($T\to 0$) limit yields the $Sl(2;\bR)$-duality and conformal invariant Hamiltonian density
 ${\cal H}=p$ of BB electrodynamics  \cite{BialynickiBirula:1984tx}, irrespective of the value of $\gamma$. Notice that 
\begin{equation}
{\cal H} =p \quad \Rightarrow \quad {\bf D} \cdot {\bf E} = {\bf B}\cdot{\bf H} =  {\cal H} \, ,  
\end{equation}
which implies (\ref{conf-inv}) and hence conformal invariance. It also implies that the attempt to find a Lagrangian density  by taking the Legendre 
transform of ${\cal H}({\bf D}, {\bf B})$ with respect to ${\bf D}$ fails, since  ${\bf D} \cdot {\bf E} -{\cal H} \equiv 0$ \cite{BialynickiBirula:1984tx}. Thus, the 
Lorentz invariance of BB electrodynamics cannot be made manifest in this way, although it can be made manifest in other 
ways \cite{BialynickiBirula:1992qj,Mezincescu:2019vxk}.  

The weak-field ($T\to\infty$) limit of (\ref{newHam}) yields the Hamiltonian density 
 \begin{equation}\label{weak-field}
{\cal H}= (\ch\gamma)s - (\sh\gamma)\sqrt{s^2-p^2} \, , 
\end{equation}
where we use  `sh'  for `sinh' and `ch' for `cosh'.   An equivalent expression (since $s^2-p^2= \xi^2+\eta^2$) is
\begin{eqnarray}\label{HamDB}
{\cal H} &= & \frac12 (\ch\gamma)( |{\bf D}|^2 +|{\bf B}|^2) \\
&& -\, \frac12(\sh\gamma)\sqrt{ (|{\bf D}|^2 -|{\bf B}|^2)^2 + 4({\bf D}\cdot{\bf B})^2} \, . \nonumber 
\end{eqnarray}
The Maxwell Hamiltonian density is recovered for $\gamma=0$, so we have now found the promised one-parameter ``ModMax'' extension of 
Maxwell's electrodynamics. For any value of $\gamma$, its  Hamiltonian density satisfies the condition 
(\ref{conf-inv}) required for conformal invariance. 

There may be other nonlinear Lorentz and duality invariant theories of electrodynamics corresponding to other solutions of 
(\ref{PDEsp}),  but there are no other such theories that are also conformal invariant.  To prove this we observe that when ${\cal H}$ is a function only of $(s,p)$ 
the conformal invariance condition (\ref{conf-inv}) becomes 
\begin{equation}\label{Euler}
s{\cal H}_s + p{\cal H}_p = {\cal H} \, . 
\end{equation}
This implies that ${\cal H}$ can be written as a product of $s$ with some function of the dimensionless ratio  $p/s$; it is convenient to choose 
\begin{equation}
{\cal H}(s,p) = sf(y) \, , \qquad y=\sqrt{1 - (p/s)^2}\, . 
\end{equation}
The condition on $f$ implied by (\ref{PDEsp}) is found to be 
\begin{equation}\label{fx}
(f- yf')^2 -(f')^2 =1\, , \qquad f'=\frac{\partial f}{\partial y}.
\end{equation}
Differentiating once we deduce that
\begin{equation}
f' {}' \left[ (1-y^2)f' + yf\right] =0\, . 
\end{equation}
This equation has two solutions:
\begin{equation}
(i) \  \ f=a+by\, , \qquad (ii) \quad  f= c\sqrt{1-y^2} \, , 
\end{equation}
and substitution into (\ref{fx}) shows that 
\begin{equation}
a^2-b^2 =1 \, , \qquad c^2 =1\, . 
\end{equation}
The first solution yields ModMax with $\tanh\gamma = -b/a$. The second solution yields
${\cal H} = \pm p$, which is the BB theory if we assume that ${\cal H}$ is non-negative.  Thus, the 
only (positive energy) conformal and duality invariant electrodynamics theories are BB electrodynamics and the new
family of ModMax theories, with Maxwell electrodynamics as the special free-field case. 

For the ModMax Hamiltonian density (\ref{HamDB}), it is convenient to define an angular variable $\Theta$ by 
\begin{equation}\label{defTh} 
\left(\xi,\eta\right) = \sqrt{\eta^2+\xi^2}\, \left(\cos\Theta,\sin\Theta\right)\, ,
\end{equation} 
where  $(\xi,\eta)$ are the rotation scalars of  (\ref{newrot});  notice that $\Theta \to \Theta +2\alpha$ 
under the duality transformation \hbox{$({\bf D}+i{\bf B}) \to e^{i\alpha}({\bf D}+i{\bf B})$}.  This allows us to write
the ModMax constitutive relations in the form
\begin{equation}\label{DefE1}
{\bf E}= \mathscr{A}_- {\bf D} -  \mathscr{C} {\bf B}\, , \qquad
{\bf H} = \mathscr{A}_+ {\bf B} -\mathscr{C} {\bf D} \, , 
\end{equation}
where 
\begin{equation}\label{coeffs}
\mathscr{A}_\pm = \ch\gamma  \pm (\sh\gamma)\cos\Theta\, , \quad \mathscr{C}= (\sh\gamma)\sin\Theta\, . 
\end{equation}

The constitutive relations linearize when $\Theta=\theta$, a constant, which suggests that we consider plane-wave configurations for which 
\begin{equation}\label{planew}
{\bf D}+i{\bf B} = \Re \,  [ \boldsymbol{\cal D} e^{i ({\bf k} \cdot {\bf x} - |{\bf k}| t)}] +i \,  \Re\, [ \boldsymbol{\cal B}  e^{i ({\bf k} \cdot {\bf x} - |{\bf k}| t)}] \, , 
\end{equation} 
where $(\boldsymbol{\cal D},\boldsymbol{\cal B})$ are complex 3-vector amplitudes. For such configurations, the field equations (\ref{feqs}) reduce to
\begin{equation}\label{BnE}
{\bf B} = {\bf n} \times {\bf E} \, , \qquad {\bf D} = - {\bf n} \times {\bf H}\, , 
\end{equation}
where ${\bf n} = {\bf k}/|{\bf k}|$, and substitution for $({\bf E},{\bf H})$ leads to the algebraic equations
\begin{eqnarray}\label{Falgebraic}
\boldsymbol{\cal D} &=& - {\bf n} \times [A_+\,  \boldsymbol{\cal B} 
- C\,  \boldsymbol{\cal D}]\, , \nonumber \\
\boldsymbol{\cal B} &=& \  {\bf n} \times  [A_- \, \boldsymbol{\cal D}  
- C\,  \boldsymbol{\cal B}]\, , 
\end{eqnarray}
where the constants $(A_\pm,C)$ are the coefficients (\ref{coeffs}) for $\Theta=\theta$. 
These equations imply that both $\boldsymbol{\cal D}$ and $\boldsymbol{\cal B}$ are orthogonal to ${\bf n}$, and they determine one in terms of the other; e.g. 
\begin{equation}\label{D-amp}
\boldsymbol{\cal D} = A_-^{-1} [C\,  \boldsymbol{\cal B} 
-  {\bf n} \times \boldsymbol{\cal B} ]  \, . 
\end{equation}
Using this result, one can deduce that 
\begin{equation}\label{sp}
p=A_-^{-1}|\mathbf B|^2\, , \qquad s= (\ch\gamma)p \, , 
\end{equation}
and that ${\bf D}\times{\bf B}={\bf n}p$, as expected. 
One can also deduce that 
\begin{equation}\label{f2rels}
\left(\xi,\eta\right) = (\tanh\gamma) s  \left(\cos\theta,\sin\theta\right)\, , 
\end{equation}
which is consistent with (\ref{defTh}) only if $\gamma\ge0$.

To summarise, for $\gamma>0$ the ModMax Hamiltonian equations admit exact light-velocity plane wave solutions, determined 
by the two non-zero independent complex components of $\boldsymbol{\cal B}$, which comprise the two amplitudes 
and relative phase of an elliptically polarised wave, and an irrelevant overall phase. However, the linear superposition of two solutions
is another solution only if both have the same direction ${\bf n}$ and the same value of $\theta$.

A feature of the Hamiltonian density (\ref{HamDB}) for $\gamma>0$ is that it is not a convex function of ${\bf D}$  for all values of $({\bf D},{\bf B})$, whereas convexity is essential for the Legendre transform to be involutive (see e.g. \cite{Arnold:1989}). Although all eigenvalues of the $3\times 3$ Hessian matrix are everywhere positive for $\gamma\le0$, the lowest eigenvalue for 
$\gamma>0$ is 
\begin{equation}\label{Hs}
{\cal H}_s  = {\rm ch}\gamma - ({\rm sh}\gamma) \, \frac{s}{\sqrt{s^2-p^2}}\, , 
\end{equation}
which is negative unless 
\begin{equation}\label{convexity}
s \ge ({\rm ch}\gamma) p\, .     
\end{equation}
From (\ref{sp}) we see that this  `convexity bound'  is saturated by the exact plane wave solution just discussed; the 
significance of this is best understood in terms of the Lagrangian formulation, to which we now turn. 

Equations equivalent to the combined Hamiltonian field equations (\ref{feqs}) and constitutive relations (\ref{EHdefs}) may be derived from 
the phase-space action
\begin{equation}
I[{\bf A};A_0]  = \int\! dt \! \int\! d^3x \left\{ {\bf E}\cdot{\bf D} - {\cal H}({\bf D},{\bf B}) \right\}\, , 
\end{equation}
where the independent fields are ${\bf D}$ and the potentials $(A_0, {\bf A})$ defined, up to gauge transformations, by the relations 
\begin{equation}
{\bf E} = \boldsymbol{\nabla} A_0 - \dot{\bf A}\, , \qquad {\bf B} =\boldsymbol{\nabla} \times {\bf A}\, .   
\end{equation}
Elimination of ${\bf D}$ effects the Legendre transform of ${\cal H}({\bf D},{\bf B})$ with respect to ${\bf D}$ and hence yields
an action with configuration-space Lagrangian density ${\cal L}({\bf E},{\bf B})$.

To implement this transform for the ModMax theory,  we first use the fact that
$\sqrt{s^2-p^2} = \xi \sec\Theta$ to rewrite the ModMax Hamiltonian density \eqref{weak-field}  as
\begin{equation}\label{Hamdensity}
{\cal H} = (\ch\gamma) s - (\sh\gamma) (\sec\Theta) \xi\, .  
\end{equation}
Next, we contract the first of equations (\ref{DefE1}) with ${\bf D}$ to obtain an expression for ${\bf E}\cdot {\bf D}$, and hence deduce that
\begin{equation}\label{ED-H1}
{\mathcal L}={\bf E}\cdot {\bf D} - {\cal H}  = \xi{\mathcal H}_s \, . 
\end{equation}
We should be able to rewrite this expression  in terms of the two independent Lorentz scalars:
\begin{eqnarray}\label{SandP11}
S &=& \frac12(|{\bf E}|^2 -|{\bf B}|^2)= -\frac 14 \eta^{\mu\nu}\eta^{\rho\sigma} F_{\mu\rho}F_{\nu\sigma} \, , 
\nonumber \\ 
P &=& {\bf E}\cdot {\bf B}= -\frac 18 \epsilon^{\mu\nu\rho\sigma}F_{\mu\nu}F_{\rho\sigma}\, , 
\end{eqnarray}
where $F_{\mu\nu}$ ($\mu,\nu =0, 1,2,3$) are the components of the field-strength 2-form $F=dA$ for the 1-form potential $A= dtA_0+ d{\bf x} \cdot {\bf A}$, and $\eta^{\mu\nu}$ is the Minkowski metric.  

Expressions for $(S,P)$ may be obtained from the equation for ${\bf E}$  in (\ref{DefE1}) by (i) taking the norm-squared of both sides (and then subtracting $|{\bf B}|^2$), and 
(ii) contracting both sides with ${\bf B}$. This yields
\begin{equation}\label{Bcon1} 
S= { [(\ch\gamma)- (\sh\gamma)\sec\Theta]\cal L} \, , \quad P=\tan\Theta\, {\cal L}\, , 
\end{equation}
with ${\cal L}$ given by  (\ref{ED-H1}). Substitution for $\xi$ using (\ref{f2rels}) yields ${\cal L}=0$, so 
$S=P=0$ for the plane-wave solution, as one would expect. Notice too that ${\cal L}=0$ is possible only if $\gamma\ge0$, which is therefore necessary for the existence of the plane-wave solution, as we have already seen.

From \eqref{Bcon1} and \eqref{ED-H1}, and the definition of $\mathscr{A}_-$ in (\ref{coeffs}), we find that
\begin{equation}
S^2+P^2 = \left[\mathscr{A}_- (\sec\Theta) \xi {\cal H}_s\right]^2\, . 
\end{equation}
As  $\mathscr{A}_-$ and $\xi\sec\Theta$ are  non-negative but ${\cal H}_s$ may be negative when $\gamma\ge0$, the sign of ${\cal H}_s$ 
will appear on the right hand side when we take the (positive) square root. 
We saw earlier that ${\cal H}$ is a  convex function of ${\bf D}$ when its domain is restricted by ${\cal H}_s\ge0$, so we make this choice. 
Then, since $\xi {\cal H}_s={\cal L}$,  we have
\begin{equation}\label{sqrtSP}
\sqrt{S^2+P^2} = (\ch\gamma\sec\Theta-\sh\gamma) {\cal L}\, . 
\end{equation}

Finally, by substitution for $S$ and $\sqrt{S^2+P^2}$, one may verify that 
\begin{equation}\label{final-Lag11}
{\cal L} = (\ch\gamma) S + (\sh\gamma) \sqrt{S^2+P^2}\, .  
\end{equation}
We have now found the manifestly Lorentz invariant ModMax Lagrangian density. For $\gamma\ge0$  it is a convex function of ${\bf E}$ for all values of $({\bf E},{\bf B})$. Also, and as expected, it 
has the form required for conformal invariance: ${\cal L}= Sf(P/S)$ for some function $f$ \cite{Kosyakov:2007qc}. 
This implies the following relation that will be used below: 
\begin{equation}\label{EulerL}
{\cal L}_{SS} {\cal L}_{PP} - {\cal L}_{SP}^2 =0\, . 
\end{equation}

The ModMax Hamiltonian density can be recovered from its Lagrangian density by an inverse Legendre transform, 
implemented by solving the equation ${\bf D}= \partial{\cal L}/\partial{\bf E}$ for ${\bf E}$
as a function of $({\bf D},{\bf B})$.  This equation for ${\bf D}$ may also be used to derive expressions 
for $(\xi,\eta)$ in terms of $(S,P)$, and these can be shown to imply that ${\cal H}_s\ge0$ 
when $\gamma\ge0$, and hence that the convexity bound (\ref{convexity})  is satisfied, in accord with
the general theory of the Legendre transform. In particular, Hamiltonian configurations that violate the convexity bound (which include all those for which the 
Hamiltonian field equations are not defined)  do not correspond to Lagrangian configurations. 
Hamiltonian configurations saturating the convexity bound, such as the exact plane wave solutions, 
correspond to Lagrangian configurations with $S^2+P^2=0$ for which the EL equations are not defined, 
because of the non-analyticity of ${\cal L}$ at such points. However, the Hamiltonian field equations 
remain well-defined, and we think it likely that these equations will consistently evolve incoming waves 
to outgoing waves through an interaction region, within which the EL equations provide an equivalent 
configuration-space trajectory. 

Because the ModMax Lagrangian density is non-analytic at $S^2+P^2=0$, the EL equations cannot be linearized about the vacuum. 
However, they can be linearized about solutions for which $S^2+P^2\ne0$,   such as one for which  ${\bf E}$ and ${\bf B}$ are generic uniform constants.  For the generic Lagrangian density ${\cal L}(S,P)$ the resulting linear equation was shown in  \cite{BialynickiBirula:1984tx} to have plane-wave solutions with a wave $4$-vector $k=(\omega,{\bf k})$  satisfying 
 \begin{equation}\label{disp}
k^2 = (i_k F)^2 \lambda_\pm \qquad  (i_kF)_\nu := k^\mu F_{\mu\nu}\, , 
\end{equation}
where $F$ is the background 2-form field-strength and $\lambda_\pm$ are ``birefringence indices''.  
The formula in \cite{BialynickiBirula:1984tx}  for these indices may be simplified for 
conformal theories by use of (\ref{EulerL}), and the result is
\begin{equation}
\lambda_-=0\, , \qquad \lambda_+ = \frac{{\cal L}_{SS} + {\cal L}_{PP}}{{\cal L}_S+ 2(P{\cal L}_{SP} -S{\cal L}_{PP})} \, . 
\end{equation}
Thus, a consequence of conformal invariance (albeit  broken by the constant background solution of the full field equations) is that there 
is always one mode with the free-field dispersion relation 
$\omega^2=|{\bf k}|^2$. 

For the ModMax theory we find that 
\begin{equation}
\lambda_+= \frac{\tanh\gamma}{\sqrt{S^2+P^2} - (\tanh\gamma)S}\, .  
\end{equation}
As shown in \cite{BialynickiBirula:1984tx}, the background mimics an optical medium in constant uniform motion. 
In the rest-frame of this medium, for which ${\bf E}$ and ${\bf B}$ are (anti)parallel, 
the  $\lambda_+$-dispersion relation of (\ref{disp}) is
\begin{equation}\label{disp+}
\omega^2 = |{\bf k}|^2 (\cos^2 \varphi + e^{-2\gamma}\sin^2\varphi)\, , 
\end{equation}
where $\varphi$ is the angle between ${\bf k}$ and ${\bf B}$; there is no birefringence when  $\varphi=0$ because in this case the background preserves rotational symmetry in the plane 
defined by ${\bf k}$.  Notice that  $\omega$  is independent of the strengths of the background fields; this is presumably a  
consequence of the conformal and duality invariance of ModMax electrodynamics. Notice too that the propagation is  superluminal for $\varphi=\pi/2$ when $\gamma<0$, 
as claimed earlier, whereas it is never superluminal for $\gamma\ge0$. 

In view of this birefringence prediction, it seems likely that compatibility with current observations involving light propagating through slowly-varying magnetic fields (terrestrial or astronomical) will put a stringent upper bound on the ModMax coupling constant $\gamma$. As the magnitude of the birefringence is independent of the (non-zero constant) strength of the background field, it is qualitatively different from that predicted by effective classical nonlinear theories of electrodynamics incorporating QED corrections, 
for which the magnitude  increases with increasing magnetic field strength (see e.g. \cite{Rebhan:2017zdx,Karbstein:2019oej}). However this QED prediction has yet to be confirmed experimentally \cite{Ejlli:2020yhk} and it is possible that an initial experimental confirmation of birefringence could have a ModMax interpretation.

\smallskip

\noindent\textbf{Acknowledgements}: We are grateful to an anonymous referee of an earlier version of this paper for suggesting  that we investigate the birefringence effect.
This work was partially supported by STFC consolidated grant ST/P000681/1 (PKT), by the Spanish MINECO/FEDER (ERDF EU)  grant PGC2018-095205-B-I00 (IB), by the Basque Government Grant IT-979-16 (IB) and by the Basque Country University program UFI 11/55 (IB).

\end{document}